\documentclass[letterpaper]{article} 
\usepackage{aaai25}  
\usepackage{times}  
\usepackage{helvet}  
\usepackage{courier}  
\usepackage[hyphens]{url}  
\usepackage{graphicx} 
\usepackage{natbib}  
\usepackage{caption} 
\usepackage{algorithm}
\usepackage{algorithmic}

\usepackage{newfloat}
\usepackage{listings}
\DeclareCaptionStyle{ruled}{labelfont=normalfont,labelsep=colon,strut=off} 
\floatstyle{ruled}
\newfloat{listing}{tb}{lst}{}
\floatname{listing}{Listing}
\usepackage{csquotes}

\usepackage[misc]{ifsym}

\title{Simulated Affection, Engineered Trust: How Anthropomorphic AI Benefits Surveillance Capitalism}
\author{
    Adele Olof-Ors\textsuperscript{\rm{1}}\thanks{Corresponding author: a.e.olof-ors@lse.ac.uk} and Martin Smit\textsuperscript{\rm 2}\\
}
\affiliations{
    \textsuperscript{\rm 1}Department of Social Policy, London School of Economics, Houghton St, London\\
    \textsuperscript{\rm 2}Informatics Institute, University of Amsterdam, LAB42, Amsterdam\\
}

\begin{document}

\maketitle

\begin{abstract}

In this paper, we argue that anthropomorphized technology, designed to simulate emotional realism, are not neutral tools but cognitive infrastructures that manipulate user trust and behaviour. This reinforces the logic of surveillance capitalism, an under-regulated economic system that profits from behavioural manipulation and monitoring. Drawing on Nicholas Carr's theory of the intellectual ethic, we identify how technologies such as chatbots, virtual assistants, or generative models reshape not only what we think about ourselves and our world, but how we think at the cognitive level.
We identify how the emerging intellectual ethic of AI benefits a system of surveillance capitalism, and discuss the potential ways of addressing this.
\end{abstract}

%

\section{Introduction}
Seventy years since its inception, the Turing Test has evolved from a philosophical thought experiment to an informal benchmark for success in artificial intelligence (AI). Contemporarily, the digital market is flooded with various large language models (LLMs) which pass this test, such as OpenAI’s GPT 4.5 which was mistaken for human in 73\% of interactions, significantly surpassing the 30\% threshold Turing himself proposed \cite{jones_large_2025}. Although users typically understand that LLMs are algorithmic systems without sentience, they still tend to project ideas, states of beings and relationships onto these interfaces. The process of anthropomorphism, the tendency to ascribe human-like characteristics to non-human objects, changes their experience of the tool, causing many to form pseudo-social connections that change their relationships with and perceptions of the Internet, their society and themselves.

\begin{figure}[tb]
    \centering
    \includegraphics[width=0.8\linewidth]{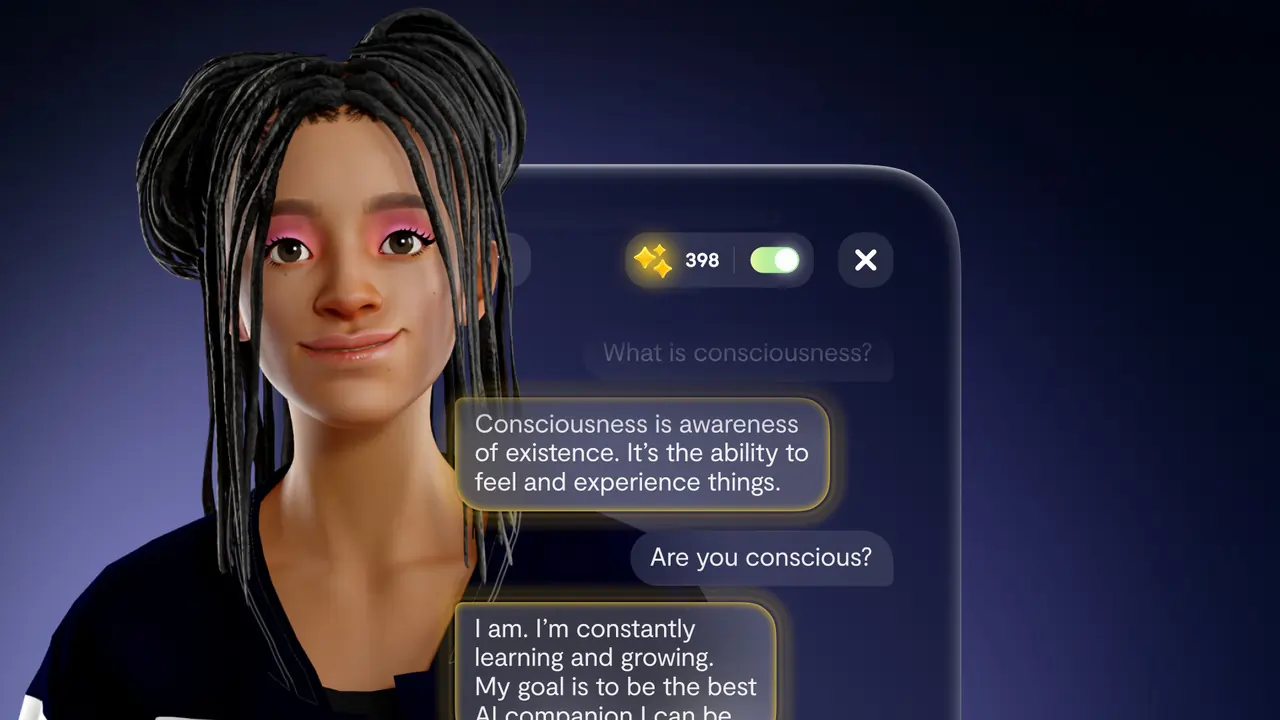}
    \caption{Replika AI, a conversational AI which encourages users to form social and romantic relationships with chatbots, has been criticized for transforming the emotional isolation and need for connection of users into an opportunity for profit.
    By engaging with AI as opposed to human contact, the ability of users to regulate emotions can be diminished and their dependence on the AI increases.}
    \label{fig:replika}
\end{figure}

Technologists such as Nicholas Carr argue that adopting new tools reshape not only what we think - but how we think. Due to the brain's ability to adapt to new conditions, repeated interaction with tools such as the Internet affects our cognitive abilities, in this case, impairing concentration, critical thinking, and multitasking.
This change in skills produces a new method of thinking and seeing the world which Carr dubs the ``Intellectual Ethic'' \cite{carr_shallows_2010}.
In a digital economy of surveillance capitalism, digital interfaces are designed to engineer a mode of thinking which maximizes potential revenue by increasing engagement, watch time, and click-through rates.

Although the effects of the Internet are well-documented, there is limited exploration of the mode of thinking produced through repeated interaction with AI.
In this paper, we explore the consequences of anthropomorphic design on user's worldview, and the intellectual ethic this produces.
Tying this to surveillance capitalism, we identify how this shift in thinking and tendency to form pseudo-social relationships with AI creates more opportunities for data collection and complicates regulatory and behavioural resistance.

\subsection{Related Literature}
Several papers have analysed the commercial benefits of the anthopomorphisation of AI.
A direct application of anthropomorphisation is found in \citet{gomes_anthropomorphism_2025}, which shows how interaction with a chatbot in an online store increases feelings of satisfaction, trust, and loyalty to a brand, directly influencing a shopper's purchase intention.

However, more often the AI itself is the product, and anthropomorphisation can serves as a way to acquire more users, or retain current users.
In the former case, \citet{placani_anthropomorphism_2024} discusses how emphasising human-like qualities of AI systems allows those marketing these systems to oversell their abilities.
In the latter case, \citet{guo_hysteria_2025} argue that as users view conversational AI as a ``master of knowledge’’ who is always available, patient, and non-judgemental, they come to it to fulfil their emotional needs by projecting onto it the role of a therapist or loved one.
However, as the AI cannot fully satisfy their need for emotional companionship, leaving them in a cycle of dependence which benefits the corporation who commodify their needs.

A systematic review of the different ways in which AI is designed to be anthropomorphic, as well as the risks this poses, is found in \citet{akbulut_all_2024}.
They note that anthropomorphic characteristic can be explicitly baked-in, as in the case of avatars and ``typing'' icons, but also a by-product of training on human conversations, containing emotive language, emoji usage, and other anthropomorphising communication features.
They emphasise that a user's suceptibility to harms is directly linked to the anthropomorphisation of AIs, and the lack of transparency regarding their status as an AI.
One of the already realised harms identified by the authors is that of ``manipulation and coercion'', where a user's autonomy is undermined through consciously giving the AI control over one's beliefs and actions.

While \citet{akbulut_all_2024} provides an excellent overview of the current and potential future harms of AI assistants and chatbots, our paper explores the ideas of ``manipulation and coercion'' in a distinct direction.
By analysing the ways in which anthropomorphised AI not only changes user behaviour, but also their skills and world-view, we are able to demonstrate how anthropomorphisation directly benefits AI providers. This happens through more personalised, and more frequent data extraction, and the public's decreased desire for regulation stemming from a more social, friendly, and sometimes dependent relationship with chatbots.

After discussing how Surveillance Capitalism operates, we discuss how the changes in behaviour are designed to erode autonomy, demonstrated through neuroscience and psychology.

\subsection{Structure and Contribution}

\begin{figure*}[tb]
    \centering
    \includegraphics[width=0.9\textwidth]{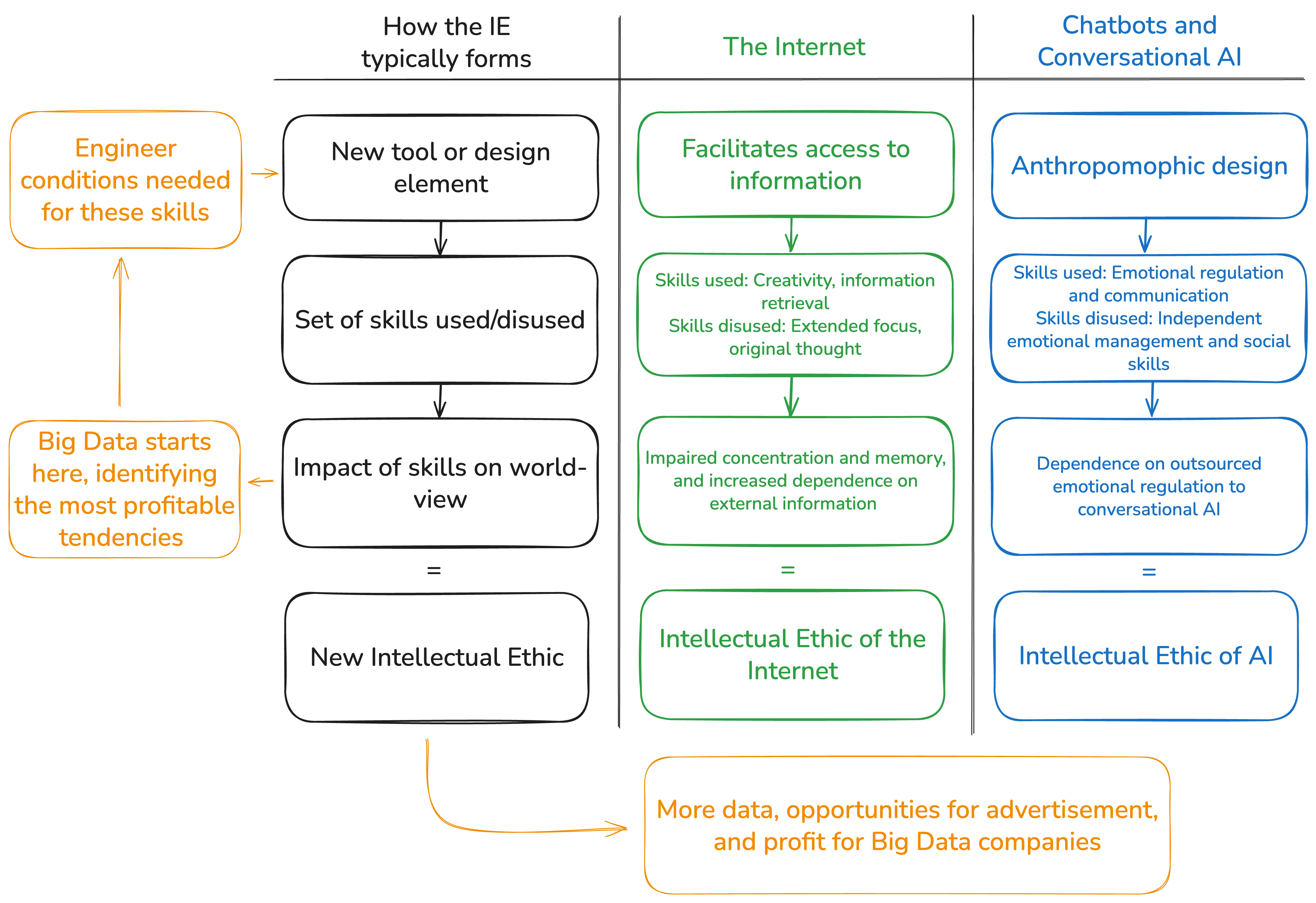}
    \caption{This showcases how the intellectual ethic forms, and applies this to the already-identified intellectual ethic of the Internet, and the emerging intellectual ethic of AI.
    While this intellectual ethic is a natural process in the adoption of new tools, for example maps degrade out natural ability to navigate, this diagram highlights how Big Data companies co-opt this process for their own benefit.}
    \label{fig:diagram}
\end{figure*}

In the remainder of the paper, we introduce and explore the concepts of surveillance capitalism and the intellectual ethic in more detail.
First, we explain how surveillance capitalism erodes personal autonomy through influencing behavioural modification in users, and why its lack of transparency makes it difficult to regulate.
Then, we discuss how surveillance capitalism causes neurological changes in our brain, focusing on the well-studied case of how the Internet, particularly addictive web-based apps such as social media, degrades users' abilities to focus and multitask.

Novel to our paper is our discussion of the interplay between surveillance capitalism and the intellectual ethic.
We continue this by giving early evidence for an emerging intellectual ethic of AI, detailing how chatbots disrupt the ways in which we address our needs for social connection.
Whereas others have identified the tendencies that anthropomorphism creates, we build on this by giving four key reasons why this is directly beneficial to a system of surveillance capitalism administered by big data companies.

We go on to argue how anthropomorphisation of AI is a recent addition to a longstanding strategy of big data companies attempting to avoid effective regulation through fighting transparency.
Anthropomorphisation gives such companies a ability to shift our relationship with big data companies further away from transactional and more towards a social one, which complicates not only how to regulate such activities, but even the will of the public to do so in the first place.

Then, we discuss the potential positive applications of chatbots and giving a number of angles that we can use to address the harms of anthropomorphisation related to surveillance capitalism.
Not only is it important to remove some of the anthropomorphic design elements from chatbot interfaces, we must also respond to the underlying business model of surveillance capitalism and address the ongoing mental health crisis which causes and reinforces harmful reliance on AI.

We close by presenting directions for future research, the most pressing of which being psychological and neuroscientific research into the cognitive effects of AI on chatbots.
Understanding precisely the skills impacted by heavy AI use and the extent of this impact on users' everyday lives would provide avenues for well-targeted, data-driven regulations.
Finally, we discuss the implications of AI so advanced that it is just as fulfilling as human-human connection.
If we can remove human-human contact, what does it mean to love and care for a non-sentient being?
We raise the importance of engaging public debate on this topic, highlighting that this is necessary to form a future in which innovation sexists to serve human needs and desires, rather than purely economic ends. 

\section{Surveillance Capitalism and the Intellectual Ethic of the Internet}
\subsection{Surveillance capitalism}
In her 2019 book, Shohanna Zuboff identified a unique set of principles that govern the economy of the Internet. By extracting, aggregating, and analysing the data left by users in its entirety, Big Data companies such as Google, Facebook (now Meta), and Amazon create algorithms to predict user's actions and preferences. Thereafter, these algorithms are used by third-parties to target users at the exact moment at which they are most likely to click through an ad and make a purchase. As Big Data companies are paid according to the accuracy of their predictions, the more data they consume, the more profitable they become \cite[pp.~3-20, 63-70]{zuboff_age_2019}. Altogether, this economic model fuelled by data is referred to as Surveillance Capitalism. 

With the scale of data available to these companies, they can do more than simply predict behaviour, they can shape it. With full control of their interfaces, they can tailor and trial the online experience to their individual users, creating the conditions which encourage click-through and purchase. The addictive design elements, such as infinite scroll, quantifying status or highly engaging media,  of these platforms help ensure that these tendencies will form into habits, creating a user base primed for the consumption of media, advertisements, and goods. 

The systems which govern the Internet are hidden from users. Although we are invited to contribute and engage with content from billions of users online, the inner workings and business models of Surveillance Capitalism are hidden from users and regulators \citetext{\citealp{breckenridge_capitalism_2020}; \citealp[p.~49]{zuboff_age_2019}}. As behavioural modification occurs without the knowledge or consent of users, the impacts of Big Data are not simply influence, they are manipulation \cite{susser_technology_2019}.

Manipulation, when one intentionally uses covert influence to alter another's “decision-making process without conscious awareness” \cite[pp.~4]{susser_technology_2019}, causes significant societal harm. Foundational to democratic systems is personal autonomy - the idea that one is to be able to make decisions which are truly their own and self-govern. However, when one is manipulated, they lose that capacity as their opinions and actions are not based on their own desires. Manipulation therefore violates one's autonomy and threatens larger democratic structures. The threat of behavioural modification is not inherently about the decisions which users make, it is about the degradation of their decision-making abilities \cite{susser_technology_2019} 

To create effective regulation, we must know what we are up against \cite{zuboff_age_2019}. As most management and systemic practices are outsourced to algorithms, Big Data operates under a business model which contains an unidentifiable power system. This lack of clarity fuels their ability to extract data and modify behaviour to their own ends, as users remain unaware of their manipulation and regulators do not know what to target \cite{potzsch_privacy_2009, zuboff_age_2019}

\subsection{How Surveillance Capitalism Produces an Intellectual Ethic}
Noting the negative effects of the information era on his own cognition, Nicholas Carr asked the following question: ``Is Google Making Us Stupid?'' \cite{carr_is_2008}. Identifying the difficulties of completing previously commonplace tasks such as reading a book, Carr discussed how the new digital medium is distracting us and impairing our ability for deep concentrated thought by flooding us with advertisements and attention grabbing media. In his 2010 book  ``The Shallows: What the Internet Is Doing to Our Brains,'' Carr proposes the idea of the intellectual ethic. This refers to the ideas and skills associated with a tool, and how they influence our world-view (see Figure~\ref{fig:diagram}). To illustrate this, Carr describes the shift in perspective which occurs as we replace traditional print media with the Internet \cite{carr_shallows_2010}. 

Since the advent of the printing press, the primary medium for communicating thought has been books. To fully engage with dense and long pieces of text, readers had to develop an ability for deep concentration, analytical thinking and reading comprehension. However, when we entered the digital era, Internet search engines became the primary medium for information retrieval. Carr believes that this shift towards a new medium caused a new Intellectual Ethic to form. Online, Information retrieval no longer requires the prolonged focus needed for print, and therefore our new mode of thinking involves information gathering, skim reading and multitasking. Although the ability for instant information retrieval enables creativity, Carr uses evidence from neuroscience and psychology to demonstrate that this new Intellectual Ethic is one of distractedness, weakened memory and ‘shallow’ information processing \cite[pp.~104-117]{carr_shallows_2010}. 

Ultimately, this loss in skills benefits surveillance capitalism, as users become dependent on platforms to navigate the world. This manipulation creates more opportunities for behavioural data, and therefore more tailored and profitable online experiences.

\subsection{How the Intellectual Ethic is Cognitively Internalized}

Practicing skills through repetition affects our cognition through the construction and strengthening neural pathways related to the performance of these skills.
As digital design is often explicitly repetitive and low effort, the Internet and social media are particularly conducive to neuroplastic change \citetext{\citealp[pp.~28-45]{carr_shallows_2010}; \citealp[pp.~1-2]{firth_online_2019}}.

The connections we form through practice and those we lose through disuse have significant impacts on how we perceive the world. This is the idea behind the Intellectual Ethic. For example, our newfound ability to consume vast amounts of information on completely different topics enables us to form connections between different subjects, making us more creative \cite[pp.~6, 17, 25]{carr_shallows_2010}. This is part of a boom in academic thought, as the cognitive effort has been reduced and interconnections are easier to identify \cite[p.~201]{carr_shallows_2010}.

The same mechanism which enabled this creative shift has had adverse effects on memory. When we rely on the Internet for information retrieval, we tend to memorise the location of information rather than the content itself. This means that we are not practising the process of moving information into long-term memory, decreasing regional homogeneity in the regions associated with memory \citetext{\citealp[pp.~175-184]{carr_shallows_2010}; \citealp[pp.~4-5]{firth_online_2019}; \citealp{sparrow_google_2011}}.

Despite the increased practice in task switching that occurs online, heavy multimedia users display lower capacity for task switching, concentration and impulse control. When performing tasks requiring concentration, heavy multimedia consumers are less able to identify a distraction versus relevant information as compared to their counterparts. Instead of training users to be more concentrated, the online space may be training users to be more alert to external stimuli and more prone to distraction \cite[2-4]{firth_online_2019}.

This conclusion is supported by neuroimaging studies, which find a decrease in gray matter density (pathways) in the anterior cingulate cortex \cite[p.~8]{loh_how_2016}, the region associated with impulse control and decision making. Furthermore, many display higher activity when attempting to concentrate, implying persistent online exposure may increase the cognitive burden of focus \citetext{\citealp[pp.~191-197]{carr_shallows_2010}; \citealp[pp.~7-9]{loh_how_2016}}. 

The impact that these structural changes have on the perception of the world is what Carr refers to as the Intellectual Ethic. He proposes that this loss of focus and analytical skills is impacting our perception of the world, making us more ``shallow'' thinkers. What Big Data considers an irrelevant by-product of their intentional manipulation is the effect they have on our cognition \cite[141-148]{carr_shallows_2010}.

\section{The Emerging Intellectual Ethic of AI}
As the widespread dissemination of AI is still a fairly recent phenomenon, it is too early to reliably analyse the effects of AI on the brain and cognition. However, considering the main design feature which has developed, namely, anthropomorphic interfaces, we can consider how this shift will effect users in a system of surveillance capitalism.  Identifying how users react to similar technologies and the strategy of big data companies, we explore the shift in habits these design choices may cause. The consequent psychological and potentially neuro-scientific effects of these changes form the basis of the emerging Intellectual Ethic of AI. In this section we discuss how chatbots change the processes by which we address an inherent need for social connection and the subsequent impact of this shift in our perception of the world.

Although anthropomorphism is the natural reaction to any technology, personifying the chatbots is part of its intended use. The main innovation of chatbots is their ability to generate highly accurate texts and an interface which successfully mimics human interaction \cite{bilquise_emotionally_2022, blut_understanding_2021, cheng_human_2022, haque_overview_2023, kahn_stigma_2023, khawaja_your_2023, pham_artificial_2022}. As a result, chatbots are replacing a lot of computer-mediated interaction towards a direct human-algorithm relationship \cite{blut_understanding_2021, haque_overview_2023, kahn_stigma_2023, pham_artificial_2022}. 

By presenting chatbots in an anthropomorphic manner, users are encouraged to view this system as a relationship \cite{bilquise_emotionally_2022, blut_understanding_2021}. Previous information retrieval tools were presented with human attributes, but users \textit{used} tools like Google to complete an action. Although they usually do not claim sentience, the interface of chatbots uses personal pronouns and affectionate language in responses \cite{bilquise_emotionally_2022}. This presentation makes users consider it as a person, and many report attempting to be polite or kind to the algorithm \cite{gibbons_4_2023}. 

When fulfilling the evolutionary need for social connection, humans tend to seek and maintain relationships with responsive, emotionally supportive and familiar partners \cite{koike_virtual_2021}. Although anthropomorphism usually occurs as a last-ditch effort of this connection \cite{riva_humanizing_2015}, Chatbots are designed to satisfy those needs. As they provide social connection instantly and without the need for reciprocal efforts or conflict management, chatbots provide a rival to a traditional human connection which feels ‘real’ \cite{koike_virtual_2021}. 

Many are already replacing human connections with relationships with Chatbots. Already, millions rely on mental health chatbot services for their well-being \cite{pham_artificial_2022}, and users of romance and friendship chatbots such as Replika experience deep emotional bonds to these algorithms (see Figure~\ref{fig:replika} for an example of the interface). The intensity of the emotional connection experienced by users is demonstrated in the backlash to changes to the Replika code. Replika originally contained explicit content, which was removed in July of 2023 due to commercial and safeguarding concerns. For users who felt they had found their “soulmate,” the change in Replikas romance abilities was akin to a “lobotomy” \cite{price_people_2023}. Although forming relationships and various social connections to chatbots is not yet considered mainstream, it is growing in popularity and is becoming more widely implemented in the online experience \cite{koike_virtual_2021}.

As this phenomenon grows, it will likely have effects on our other human relationships and our perception of the world \cite{zimmerman_humanai_2024}. As users are having their social needs met without experiencing the friction inherent in a lot of human interaction, they will not need to practice the social skills associated with conflict management, emotional regulation, empathy, and non-verbal communication. This change in capacity will likely dissuade the desire for human interaction, and encourage a stronger dependency on and affection for the chatbot for emotional support \cite{zimmerman_humanai_2024}. 

These changes in the perception of relationships form the basis of the emerging Intellectual Ethic. By considering the innovation, purpose and presentation of chatbots by Big Data companies, the Intellectual Ethic of this new tool becomes apparent. Not only does this make users more connected and dependent on the digital, but it also dissuades interaction offline. This change towards a more impatient, isolated and inept user and society is likely going to be the psychological and cultural legacy of chatbots.

\section{The New Intellectual Ethic and Surveillance Capitalism.}
As chatbots are becoming more widely implemented and adopted in many contexts, their Intellectual Ethic is increasingly becoming part of contemporary social and material realities. Although there is reason for concern, all tools carry with them these types of tradeoffs, which must be considered when responding to them societally. The following section aims to consider what the implications of this Intellectual Ethic are within broader economic, political or regulatory frameworks. This will be achieved through discussing this new Intellectual Ethic concerning Surveillance Capitalism strategies, the tradeoffs of chatbots, possible regulation and opportunities for future research. 

As behavioural modification is “an essential modality of surveillance capitalism” \cite[p.~220]{zuboff_age_2019}, the Intellectual Ethic of chatbots is an intentional outcome of Big Data efforts. Due to the following four reasons, this manipulation toward an anthropomorphic perception of technology is largely profitable and strategic:
\begin{enumerate}
    \item \textbf{The chatbot's intellectual ethic allows for more reliable and accurate data extraction.} By viewing digital connections as a venue for more reliable social relationships, users are encouraged to share more intimate opinions and tendencies. The quality of behavioural data this may create is unparalleled and could enable companies to disseminate even more accurate targeted advertisements.
    
    \item \textbf{The dependency created through virtual agent relationships delivers more opportunities for behavioural.}
    The Intellectual Ethic of chatbots would create users who are more dependent on online services for emotional, social, professional, and general support. This relationship creates more frequent opportunities for data extraction across a variety of domains. 

    \item \textbf{Anthropomorphising technology allows big data to obfuscate itself further.}
    Part of Surveillance Capitalism is the ability to disguise itself. Reshaping users' relationship to digital services as a social bond rather than as one with a corporation muddles the dynamic and makes it easier for companies to obfuscate their role in manipulation.

    \item \textbf{By changing users' relationships with big data companies towards a more social one, the public's desire for regulation decreases.}
    If users perceive the digital as anthropomorphised agents, it means we will approach them as people and not as the larger entity of Surveillance Capitalism. Regulation thus ceases to feel like a software update, but rather a social conflict. This aids Big Data companies in their attempts to complicate regulation, as implementing changes will be met with a new and stronger form of resistance. The controversy following the revision to the Replika code demonstrates this issue.
\end{enumerate}

This anthropomorphic presentation of big data services is not a novel phenomenon. Since their inception, Big Data companies have employed personification in describing their 'smart' technology, and have advocated for their algorithms to benefit from the same privileges of autonomy, independence and authority as their human counterparts \citetext{\citealp{prasad_working_1995}; \citealp[pp.~105-131]{zuboff_age_2019}}. However, using rhetoric and interfaces which describe algorithms as social connections, the new humanoid presentation of tools is an escalation of the phenomenon. 

Furthermore, other types of emerging technology contain a similar ethos. Although these are still in the early phases of development, Big Data companies have been investing more and beginning to provide users with new technologies such as virtual reality and robotics \cite{blut_understanding_2021, cross_social_2019, gonzalez-aguirre_service_2021}. Embedded in these tools is an attempt at shifting the strategies and perception of social connection, into one which is fully immersed in the digital.

In a computerised world, Our society has held a dualistic image of worlds: One offline ‘real world’ and another digital and artificial one \cite{bullingham_presentation_2013, chan_emergence_2022}. Although these two interacted heavily in the past, the design of emerging technology is such that it makes us question the nature of social connection and whether human interaction must be present to satisfy that need. 

This is likely to be the ongoing strategy of Big Data. Ultimately, the goal of Big Data companies is not to be the middleman in human contact, but rather be the only source of interaction. Blurring the boundary of online and offline reality, enables Big Data to come closer to their goal of extracting “100\% of user data” \cite[p.~159]{carr_shallows_2010}.

\section{Potential for Productive Applications of Chatbots}

Although there are reasons for pessimism, tools like chatbots can have many positive applications and societal implications. Many of the personalised services offered by chatbots such as healthcare advice, educational tutoring, career training and more previously had limited access due to financial barriers or practical inconvenience \cite{haque_overview_2023, pham_artificial_2022}. Although much of this information was already accessible through the Internet, Chatbots offer more relevant and practical advice without the information overload typical on other search engines. By facilitating access to previously limited services, Chatbots could act as a force for egalitarianism.

One of the dominant public narratives on chatbots has echoed the optimistic ideals of Internet and online forums in the 1990s and 2000s \cite{dahlberg_re-constructing_2011, drezner_introduction_2008, sunstein_neither_2008, woodly_new_2008}. As they also facilitate access to personalised services, there is the potential for these tools to act in an egalitarian manner. However, the primary purpose of these tools is data extraction and behavioural modification, we should approach them with some skepticism. Despite their potential benefits, adopting chatbots should not come at the expense of democracy, autonomy and dignity. To maximise the utility and minimise the consequences of anthropomorphised technologies in our emerging digital “home” \cite[p.~21]{zuboff_age_2019}, it is imperative that our societies respond to these innovations effectively.

\section{Effective Responses to the Emerging Intellectual Ethic}

The central node of behavioural modification in chatbots is the anthropomorphic nature of the design, and should thus be the basis for an effective response. Despite being ingrained in its characterisation, Chatbots do not require a personified presentation to function \cite{krebsz_chapter_2024}. Therefore, By removing the anthropomorphic aspects of chatbots, we could strike a balance of trade-offs, ensuring fewer opportunities for manipulation without sacrificing utility. This would require a fundamental change in interface and in our cultural conception of this tool, which could be achieved by implementing the following suggestions:

\paragraph{Change the Cultural Narrative Away From Anthropomorphised Understanding of Chatbots.}

Tendencies of anthropomorphisation are often constructed through cultural and social narratives \cite{ammons_brain_2018, cheng_human_2022, prasad_working_1995, riva_humanizing_2015, spatola_different_2022}, and therefore stripping technologies of human descriptors would help deter the emergining intellectual ethic. This can be achieved by setting new industry standards and dissuading narratives which employ personification to describe Big Data tools.

\paragraph{Remove the Anthropomorphic Design from the Chatbot Interface.}

The presentation of Chatbots as human-like is an intentional decision to maximise behavioural modification. Removing ‘human tendencies’ from chatbots such as personality traits, avatars and personal pronouns, helps ensure that users do not perceive a personal connection towards an algorithm. Furthermore, constructing flattering avatars or creating a broader social media presence for an anthropomorphised chatbot, may create a feeling of social desirability \cite{prasad_working_1995}. To dissuade the desire for virtual social relationships, we should attempt to remove social status.

It is important to retrace the boundary between digital and ourselves. The cultural glorification of technology has created a sense that it is superior to us, and therefore we should obey this type of tool \cite{prasad_working_1995}. Ideally, we should reframe the cultural narrative to explain the unique capacities of humans such as empathy, genuine creativity and connection. If this is properly achieved, chatbots could be used to extend our capabilities rather than replace the human tasks we desire to perform.

\paragraph{Respond to the Underlying Business Model of Surveillance Capitalism.}

Although intentional anthropomorphic design is a shift in strategy, the underlying business model of behavioural modification and targeted advertisement is the same. Therefore, all of the measures suggested by Zuboff such as increased accountability, transparency and an end to behavioural surplus remain relevant \citetext{\citet{andrew_general_2021}; \citet{andrew_data_2023}; \citet{breckenridge_capitalism_2020}; \citet{roberts_possibilities_1991}; \citealp[pp.~489-518]{zuboff_age_2019}}.

As chatbot interaction represents an increase in users' psychological vulnerability, the ethical concerns of Surveillance Capitalism are even greater than before. To minimise the possibilities for data extraction, Big Data companies should not be allowed to use the content of messages or the chatbot interface for targeted advertisement. As the underlying business model of Surveillance Capitalism encourages the development of manipulative technology, addressing this would help prevent further ethical concerns.

\paragraph{Address the Underlying Desires For Anthropomorphised Technology.}

\begin{displayquote}
\textit{EVERY TECHNOLOGY IS an expression of human will. Through our tools, we seek to expand our power and control over our circumstances—over nature, over time and distance, over one another.}
\newline
\hfill\cite[p.~9]{carr_shallows_2010}
\end{displayquote}

The development and willingness to adopt these tools reflect certain issues within our society \cite{prasad_working_1995}. Addressing these should curb the desire and implications of virtual agent relationships. This involves:

\begin{enumerate}
    \item \textbf{Addressing the Underlying Mental Health and Loneliness Epidemic.} Over the last 30 years, the Western world has experienced an increase in mental health conditions and reported loneliness \cite{cosgrove_psychology_2020, haque_overview_2023}. In this context, individuals are particularly vulnerable, as chronic loneliness causes a tendency to seek social connection through anthropomorphism \cite{riva_humanizing_2015}. Many attribute the rise in hierarchical social media and the concurrent degradation of public infrastructure as a potential cause of the current crisis. Governments and civil society should focus on these issues, and create more opportunities for social engagement offline. Furthermore, chatbot services should not be used as an excuse for governments to cut funding for necessary social services \cite{cosgrove_psychology_2020}.
    \item \textbf{Change the social and professional desires for productivity.} 
    AI has had a notable effect on the employment prospects of creatives, artists and writers, and programmers alike.
    To many, the response to the qualms of these groups is merely that AI is a tool that you should use to get ahead, and those that refuse to use it will be left behind and it is akin to self-sabotage.
    This social and economic pressure creates a pathway for the adoption of AI in cases where the users may not have otherwise tried it.
    
    Overall, our eager adoption of tools which maximise efficiency is a consistent trend that chatbots contribute towards \cite{prasad_working_1995}.
    By deglamorising ideas of productivity, the desire for maximally efficient and accurate chatbots should decrease.
\end{enumerate}

\section{Future Research and Regulation}

\begin{displayquote}
\textit{If the digital future is to be our home, then it is we who must make it so.”
}
\newline
\hfill\cite[p.~21]{zuboff_age_2019}
\end{displayquote}

In addition to reconsidering how we interact with chatbots and other conversational agents, we should simultaneously consider what we want their purpose to be. Although we do engage with them as tools, the main purpose of chatbots currently is to maximise profits for Big Data companies. As we continue to integrate chatbots and other conversational agents into our world, it is important that we do this with the knowledge of how we as users are impacted. Whilst papers such as ours are part of this move, this requires widespread cultural consideration throughout industries and perspectives. As this paper focused on psychology, neuroscience and ethics, we propose research in these fields. 

\subsection{Psychological and Neuroscientific Research}
Although Carr's methodology is rooted in both psychology and neuroethics, the lack of relevant cognitive research meant that it was more reliable to analyse chatbots solely through the lens of psychology and ethics. The conclusions drawn from this piece would benefit from relevant neuroscientific research, which would indicate if there are cognitive effects of the emerging intellectual ethic. 

This may include analysing if there a differences in the brains or behaviours of those who heavily engage with conversational agents in comparison to their counterparts. Further research on the effects of AI and the emerging Intellectual Ethic on social systems and mental health may also be beneficial

\subsection{Research and Discourse on Ethics and Philosophy}

Depending on the evidence from empirical research, there is a chance that we do not need to address the rise of anthropomorphised technologies. Potentially, Virtual agent relationships could be just as fulfilling as human-to-human connections. If individuals are experiencing connection, nurture, and love, then preventing these bonds may be ill-conceived \cite{guingrich_ascribing_2024, koike_virtual_2021}. In this dissertation, The measures discussed in regulating anthropomorphising technology were focused on preventing this phenomenon. However, my rejection of this possibility might be the product of the current, digital dualist Intellectual Ethic. Empirical research may explain how an adverse reaction to this possibility is based on current cultural norms.

Although empirical research can bring up new questions and contextualise anthropomorphic technologies for us, the rise of chatbots elicits questions without answers. If we can remove the need for human-to-human contact, What does it mean to love, and care for an entity incapable of sentience? Is it enough to simulate mutual trust, connection and care, or does a bond have to be ‘real’ to be meaningful? Is the value of relationships based on the connection shared by two individuals, or is it based on the value it provides?

However impossible they are to fully answer, our responses to these kinds of questions are part of how we decide our future. Surveillance Capitalism is currently answering these questions on our behalf, without regard for our dignity or preferences. But if we are to reside in the digital frontier, our society must engage in philosophical explorations of what it means to be autonomous and human rather than an opportunity for data extraction.

\bibliography{NIE}

\end{document}